\documentclass{article}

 \usepackage[preprint]{neurips_2025}

\usepackage[utf8]{inputenc} 
\usepackage[T1]{fontenc}    
\usepackage{hyperref}       
\usepackage{url}            
\usepackage{booktabs}       
\usepackage{amsfonts}       
\usepackage{nicefrac}       
\usepackage{microtype}      
\usepackage{amsmath}
\usepackage{minted}
\usepackage{tabularx}
\usepackage{ragged2e}
\usepackage{array}
\newcolumntype{L}{>{\RaggedRight\arraybackslash}X}
\newcolumntype{Y}{>{\RaggedRight\arraybackslash}p{2.2cm}}

\usepackage{listings}
\usepackage{xcolor}

\definecolor{bordercolor}{rgb}{0.2,0.2,0.2}

\lstdefinestyle{jsonstyle}{
    basicstyle=\ttfamily\footnotesize,      
    backgroundcolor=\color{white},          
    frame=single,                           
    framesep=3pt,                           
    framerule=0.5pt,                        
    rulecolor=\color{bordercolor},          
    numbers=none,
    aboveskip=6pt,
    belowskip=6pt,
    showspaces=false,
    showstringspaces=false,
    showtabs=false,
    tabsize=2,
    keywordstyle=\color{black},
    commentstyle=\color{black},
    stringstyle=\color{black},
    breaklines=true,
    breakatwhitespace=true,
    literate=
     *{0}{{{\color{black}0}}}{1}
      {1}{{{\color{black}1}}}{1}
      {2}{{{\color{black}2}}}{1}
      {3}{{{\color{black}3}}}{1}
      {4}{{{\color{black}4}}}{1}
      {5}{{{\color{black}5}}}{1}
      {6}{{{\color{black}6}}}{1}
      {7}{{{\color{black}7}}}{1}
      {8}{{{\color{black}8}}}{1}
      {9}{{{\color{black}9}}}{1}
      {:}{{{\color{black}:}}}{1}
      {,}{{{\color{black},}}}{1}
      {\{}{{{\color{black}\{}}}{1}
      {\}}{{{\color{black}\}}}}{1}
      {[}{{{\color{black}[}}}{1}
      {]}{{{\color{black}]}}}{1}
}

\usepackage{graphicx}

\makeatletter
\usepackage{xspace}
\DeclareRobustCommand\onedot{\futurelet\@let@token\@onedot}
\def\@onedot{\ifx\@let@token.\else.\null\fi\xspace}

\def\etal{\emph{et al}\onedot}
\makeatother
\usepackage{multirow}

\title{A2H: Agent-to-Human Protocol for AI Agent}

\author{%
\textbf{Zhiyuan Liang$^1$\thanks{Equal contribution.}} \quad
\textbf{Enfang Cui$^1$\footnotemark[1]} \quad
\textbf{Qian Wei$^1$} \quad
\textbf{Rui She$^1$} \quad \and
\textbf{Tianzheng Li$^1$} \quad
\textbf{Minxin Guo$^1$} \quad
\textbf{Yujun Cheng$^2$} \\
$^1$China Telecom Research Institute, China \\
$^2$School of Intelligence Science and Technology,\\
University of Science and Technology Beijing, Beijing, China \\
\texttt{\{liangzy17, cuief,  weiq12, sher, litz4, guomx3\}@chinatelecom.cn} \\
\texttt{yjcheng@ustb.edu.cn}
}

\begin{document}

\maketitle

\begin{abstract}
AI agents are increasingly deployed as autonomous systems capable of planning, tool use, and multi-agent collaboration across complex tasks. 
However, existing agent-related protocols focus on agent-to-agent interactions, leaving humans as external observers rather than integrated participants within the agent systems.
This limitation arises from the lack of a standardized mechanism for agents to discover, address, and interact with humans across heterogeneous messaging platforms.
In this paper, we propose the A2H (Agent-to-Human) protocol, a unified protocol that enables humans to be registered, discovered, and communicated with by AI agents as resolvable entities within agent systems.
A2H contributes three key components: 
(1) Human Card for registering human identities via resolvable domain names, making them discoverable to agents; 
(2) Formal Communication Schema defines when, why, and how agents contact with human;
(3) Unified Messaging Abstraction standardizes diverse communication medias and transforms complex JSON outputs into human-friendly formats. 
This work establishes a foundational protocol for integrating humans into agent ecosystems, advancing AI agents from isolated autonomous systems toward truly human-connected intelligent infrastructures.
\end{abstract}

\section{Introduction}

AI agents are evolved beyond simple chatbots into autonomous systems. They are able to independent planning, tool use, and multi-agent collaboration \cite{guo2024large}. 
As agent workflows become more complex, agents are no longer isolated tools but as core components of an AI infrastructure. This shift leads to the rapid development of agent communication protocols \cite{yang2025survey} where specialized agents work together to achieve high-level objectives.

To facilitate multi-agent coordination, various communication protocols \cite{mcp2024,ibm2024acp,google2024a2a,anp2024,cui2025agentdns} between agents are proposed. 
For instance, context-oriented frameworks like the Model Context Protocol (MCP) \cite{mcp2024} and agents.json focus on standardizing how agents access data and tools. 
Simultaneously, inter-agent protocols such as A2A \cite{google2024a2a}, ANP \cite{anp2024}, and AComP \cite{acp2025} have emerged to govern machine-to-machine dialogues, while domain-specific solutions like Agent Protocol \cite{agentPro2025} or PXP \cite{pxp2025} address task-sharing in localized environments. 
These advancements standardize the way agents discover each another, creating an efficient but closed agent ecosystem.

However, a fundamental gap remains in existing communication protocols: only agents are included in the system, while humans are treated as external observers rather than reachable participants.
As a result, an agent has no standardized way to discover a specific human when it encounters an impasse.
There are two main challenges for agents in finding and engaging humans.
First, agents lack clear rules for deciding when and why human assistance is needed.
Second, human communication is fragmented across diverse messaging platforms, such as Slack \cite{slack}, WeChat \cite{wechat}, Microsoft Teams \cite{teams}, and email. In addition, agent outputs are usually JSON-based and difficult for humans to read.
Together, these issues block seamless interaction between AI agents and humans in autonomous systems.

To bridge this disconnect between agents and humans, we propose A2H (Agent-to-Human), a protocol that integrates humans as resolvable entities within the agent ecosystem and enables hybrid agent–human collaboration.
The core idea of A2H is to move from a \textit{human-as-trigger} model to a \textit{human-as-node} architecture through the concept of the \textit{Human Card}.
By assigning humans a standardized digital identity, A2H allows agents to query a human’s availability and expertise in the same way they invoke software tools.
In addition, A2H defines clear conditions for when an agent should seek human assistance. 
Finally, A2H introduces an A2H-JSON schema that transforms complex agent outputs into concise, human-readable messages and maps human responses back into structured formats.
Our main contributions are threefold:
\begin{enumerate}
    \item We propose A2H (Agent-to-Human), the first protocol that integrates humans into the agent discovery and resolution process through the \textit{Human Card}.
    \item We design a \textit{Formal Communication Schema} that defines when, why, and how agents could seek human for help.
    \item We introduce a \textit{Unified Messaging Abstraction} that enables bidirectional translation between agent outputs and human-readable communication.
\end{enumerate}

\section{Related Work}

The proposed A2H aims to provide a unified protocol for interactions between AI agents and humans. It builds on research in agent communication protocols and human-agent interaction. 
Existing methods do not treat humans as reachable entities in agent ecosystems. 
In the following, the related agent communication protocols and human-agent interaction will be introduced.

\subsection{Agent Communication Protocol Standardization}
The rise of AI agents has created a stronger need for standardized communication protocols \cite{yang2025survey}. Recent surveys suggest that a unified protocol could greatly improve agent–tool interaction \cite{bhardwaj2025agent}. 
Several protocols are already in use which includes the Model Context Protocol (MCP), the Agent Communication Protocol (ACP), and the Agent-to-Agent Protocol (A2A) \cite{jeong2025study}.
MCP \cite{mcp2024} defines how agents communicate with external tools. It focuses on secure and scalable integration. 
A2A \cite{google2024a2a} supports message exchange between multiple agents in a structured format. 
ACP \cite{ibm2024acp} enables both asynchronous and synchronous communication between independent agents.
Also, there are some domain-specific solutions like Agent Protocol \cite{agentPro2025} or PXP \cite{pxp2025} address task-sharing in localized environments. 
AgentDNS \cite{cui2025agentdns} provides a system that enables LLM agents to discover and utilize external services by routing requests through other LLM agents. 
It establishes a naming system for LLM agents, enabling them to operate with each other.

However, these protocols center on agent–agent or agent–tool communication. They do not cover human communication channels such as email or instant messaging. 
As for AgentDNS, it depends on a centralized registry lacking native support for human participants.
This centralized structure only works for agents and does not work for humans. 
Humans cannot be registered through the same automated process. Thus, humans cannot be discovered through current agent communication protocols. 
For these reasons, a more comprehensive human-agent interaction method is needed. It is critical for truly human-connected intelligent infrastructures. 

\subsection{Human-Agent Interaction Frameworks}
Recently, several human-agent interaction frameworks have been proposed. 
For example, Blandford \etal \cite{blandford2005dicot} introduced the Distributed Cognition for Teamwork (DiCoT) framework. It studies how information is shared and processed in human-agent teams, and also analyzes how cognitive processes are shared and coordinated across members.
Sepich \etal \cite{sepich2021human} developed Human-Agent Team Game Analysis framework. This framework evaluates the roles and interactions between humans and agents in collaborative tasks. It highlights autonomy and communication as key factors for effective teamwork.
While these frameworks provide valuable insights into the design of agentic systems, they often fall short in scalability, adaptability, and real-world deployment \cite{wang2025internet}. Future extensions will need to support more complex task recipes with branching logic and preconditions \cite{wang2025internet}. It is also difficult to integrate these frameworks with diverse communication channels and modalities. Examples include natural language interfaces, multimodal interaction systems, and existing messaging platforms \cite{chen2024internet}.
In addition, the organizers have successfully conducted multiple workshops at human-computer interaction conferences \cite{zargham2024designing}. These workshops have served as a platform for proposing numerous new methods.

However, current frameworks exhibit two key limitations.  
First, these methods are not designed for agent ecosystems.
Second, they address human-agent interactions but fail to account for those between agents.

In summary, existing research offers useful insights into agent discovery, human–agent interaction, and communication protocols. However, it still lacks a unified solution for integrating humans into agent ecosystems. AgentDNS does not natively support human participants. Current human–agent interaction frameworks also ignore discovery and routing challenges. Most communication protocols focus on agent–agent or agent–tool exchanges. They provide no clear way to bridge diverse human communication channels.

\section{Method}
In this section, we present the design and implementation of the A2H protocol. 
The A2H protocol is designed to transition the role of humans in autonomous systems from passive, external triggers to active, addressable nodes. The architecture consists of three distinct layers: (1) the \textit{Human Card} for identity registration and discovery, (2) the \textit{Formal Communication Schema} for governing interaction logic, and (3) the \textit{Unified Messaging Abstraction} for cross-platform message rendering and serialization.

\begin{figure}[htbp]
    \centering
    \includegraphics[scale=0.5]{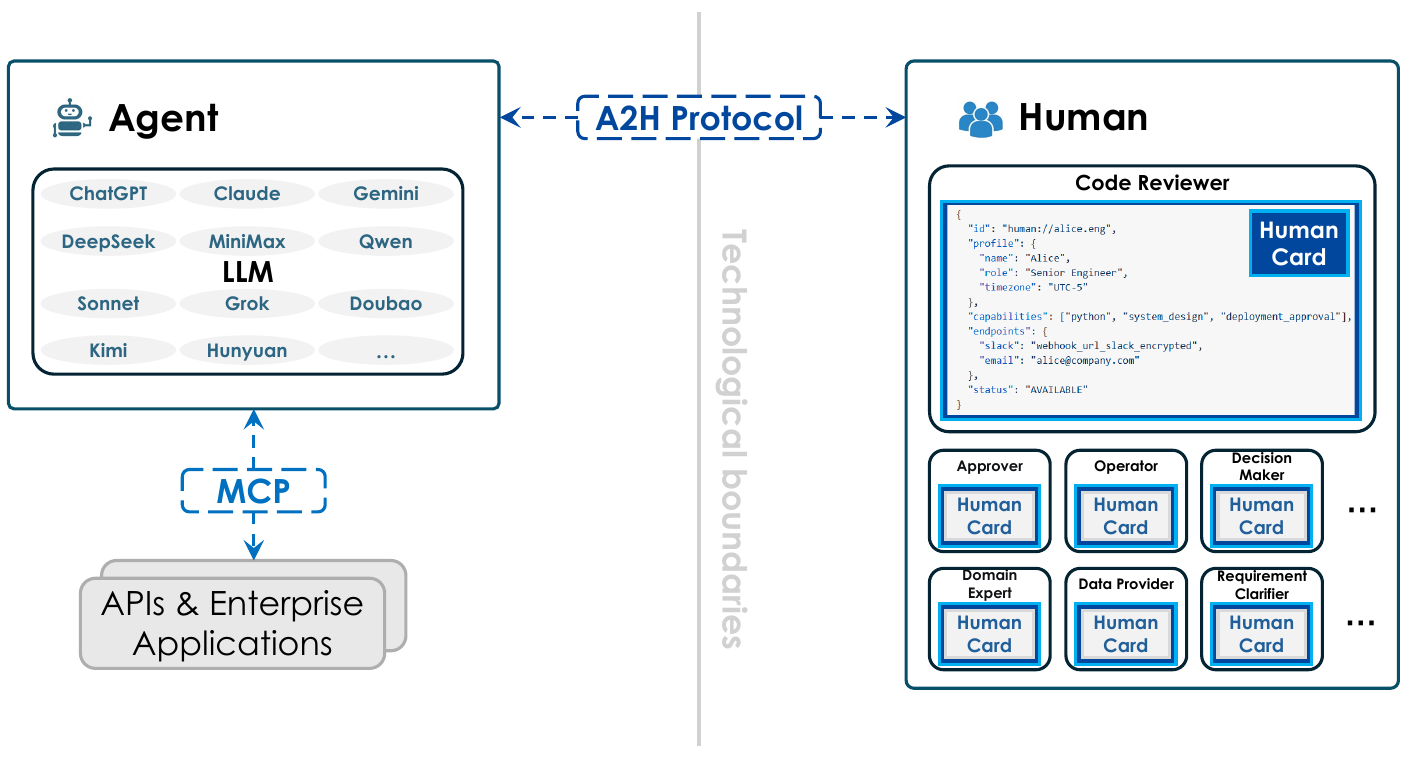}
    \caption{The A2H protocol for agent-human interaction.}
    \label{figure_arch}
\end{figure}

Figure \ref{figure_arch} illustrates the architectural overview of the A2H protocol, which establishes a bidirectional interaction layer between AI agents and humans. On the left, an AI Agent may be powered by heterogeneous large language models (e.g., ChatGPT, Claude, Gemini, DeepSeek, Qwen) and interacts with enterprise systems and external tools via MCP \cite{mcp2024} and standard APIs.
On the right, humans are modeled as discoverable entities through Human Cards, which formally encode identity, role, capabilities, availability, and preferred communication endpoints. Different operational roles (such as Approver, Operator, Decision Maker, and Domain Expert) are uniformly abstracted under the same schema.


\subsection{Human Card: Identity and Discovery Human}
To enable agents to discover humans in the same manner they discover software tools or API endpoints, we introduce the concept of the \textit{Human Card}. The Human Card serves as a structured metadata registry that encapsulates a human's identity, expertise, availability, and communication endpoints.

Formally, we define a Human Card $H$ as a tuple:
\begin{equation}
\label{eq:human_card}
    H = <\text{ID}, \mathcal{P}, \mathcal{S}, \mathcal{E}, \delta>,
\end{equation}
where:
\begin{itemize}
\item $\text{ID}$ is a unique, resolvable identifier (e.g., human://alice.eng), similar to a DNS record.
\item $\mathcal{P}$ represents profile metadata (e.g., name, role, timezone).
\item $\mathcal{S}$ is a set of semantic tags describing the human's expertise (e.g., ["python\_expert", "legal\_approver"]), allowing agents to perform semantic matching.
\item $\mathcal{E}$ denotes the set of supported communication endpoints (e.g., Slack Webhook, Email API).
\item $\delta$ represents the real-time availability status (e.g., AVAILABLE, BUSY, OFFLINE).
\end{itemize}

The Human Card is stored in a distributed registry (similar to AgentDNS \cite{cui2025agentdns} or a Key-Value store). When an agent requires assistance, it queries this registry. For example, if an agent needs code review, it executes a semantic search: $\text{Find}(h \in \text{Registry} \mid \ \text{ “code\_review”} \in h.\mathcal{S} \land h.\delta = \text{AVAILABLE})$.

A JSON representation of a Human Card is shown below:
\begin{lstlisting}{json}
{
  "id": "human://alice.eng",
  "profile": {
    "name": "Alice",
    "role": "Senior Engineer",
    "timezone": "UTC-5"
  },
  "capabilities": ["python", "system_design", "deployment_approval"],
  "endpoints": {
    "slack": "webhook_url_slack_encrypted",
    "email": "alice@company.com"
  },
  "status": "AVAILABLE"
}
\end{lstlisting}

\subsection{Formal Communication Schema}
Unlike agent-to-agent communication, which is high-frequency and low-latency, agent-to-human communication must be minimized to avoid cognitive overload. Therefore, we define a Formal Communication Schema that governs \textit{when} and \textit{why} an agent initiates contact.

We model the agent's decision process as a function $f_{decide}$ that maps the current agent state $S_t$ to an action space $A$:

\begin{equation}
    f_{decide}(S_t) \rightarrow \{\text{CONTINUE}, \ \text{HALT}, \ \text{REQUEST\_HUMAN}\}
\end{equation}

The transition to $\text{REQUEST\_HUMAN}$ is triggered by three primary conditions:
\begin{enumerate}
    \item Ambiguity Trigger ($\tau_{amb}$): When the confidence score of the agent's next planned action falls below a predefined threshold $\epsilon$.
    \begin{equation}
        \text{if} \ P(action_{next} \mid S_t) < \epsilon , \text{Trigger}\ \text{A2H}
    \end{equation}
    \item Criticality Trigger ($\tau_{crit}$): When a proposed action involves irreversible side effects (e.g., deleting a database, transferring funds). These actions are flagged in the protocol manifest as REQUIRE\_APPROVAL.
    \item Resource Exhaustion ($\tau_{res}$): When the agent enters a loop or exceeds the maximum number of reasoning steps without reaching a terminal state.
\end{enumerate}

Once triggered, the protocol generates an Intent Packet containing the \textit{Reason} (Why), the \textit{Context} (What happened), and the \textit{Required Input} (What the human needs to do).

\subsubsection{Interaction Primitives}
Once a communication trigger condition ($\tau$) is met, the agent must instantiate a specific interaction type. We categorize the communication requirements into four fundamental interaction primitives. These primitives define the semantic intent of the message and the expected structure of the human response.
Table \ref{tab:primitives} illustrates the taxonomy of these primitives within the A2H protocol.

\begin{table}[ht]
    \centering
    \caption{Taxonomy of A2H Interaction Primitives}
    \label{tab:a2h_interaction}
    \small
    \begin{tabularx}{\textwidth}{Y L L L}
        \toprule
        \textbf{Primitive Type} & \textbf{Semantics} & \textbf{Agent State Constraints} & \textbf{Expected Human Response} \\
        \midrule
        PERMISSION & 
        Authorization for high-risk actions (e.g., fund transfer, file deletion). & 
        \textbf{Hard Block:} Agent pauses execution immediately. & 
        Boolean (ALLOW / DENY) \\
        
        CLARIFICATION & 
        Resolving ambiguity when multiple valid paths exist ($P(a_1) \approx P(a_2)$). & 
        \textbf{Soft Block:} Agent pauses the current thread. & 
        Selection (OPTION\_A / OPTION\_B) \\
        
        SOLICITATION & 
        Requesting missing information necessary for the next step (e.g., API keys, files). & 
        \textbf{Soft Block:} Agent waits for specific data injection. & 
        Structured Data (String, File, JSON) \\
        
        NOTIFICATION & 
        Informational updates or task completion reports. & 
        \textbf{Non-Blocking:} Agent continues execution. & 
        None (Acknowledgment optional) \\
        \bottomrule
    \end{tabularx}
    \label{tab:primitives}
\end{table}

Formally, the interaction $I$ is defined as:
\begin{equation}
    I = \text{Type}(S_t) \times \text{Payload}(S_t),
\end{equation}
where $\text{Type} \in { \texttt{PERM}, \texttt{CLAR}, \texttt{SOLI}, \texttt{NOTI} }$.

\subsubsection{Communication Patterns}
\begin{figure}[htbp]
    \centering
    \includegraphics[scale=0.32]{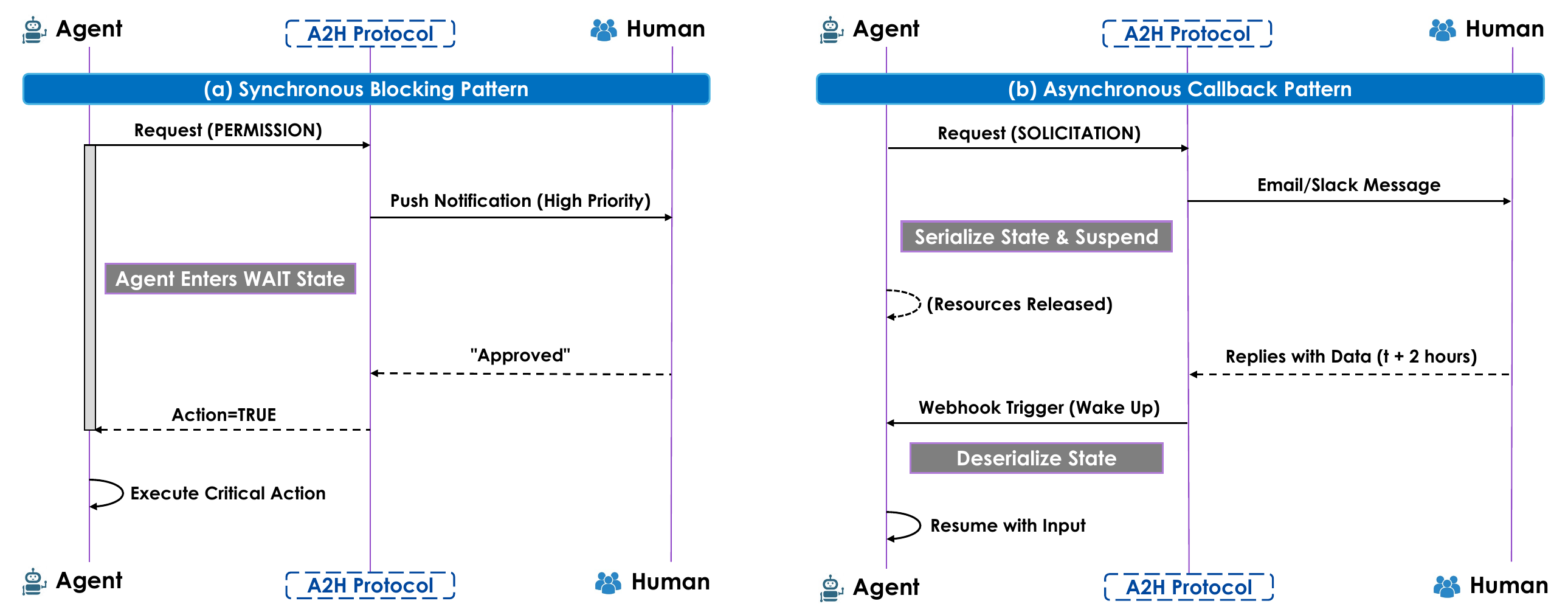}
    \caption{Sequence diagram contrasting Synchronous Blocking vs. Asynchronous Callback patterns in A2H.}
    \label{figure_communication}
\end{figure}

The A2H protocol supports different temporal flows to accommodate the disparity in processing speeds between agents (milliseconds) and humans (minutes to hours). We define two primary \textit{Communication Patterns}: Synchronous blocking and asynchronous interrupt. Figure \ref{figure_communication} illustrates the sequence flow of these two patterns.

\begin{enumerate}
    \item Synchronous Blocking (Immediate Intervention). It is used primarily for PERMISSION and urgent CLARIFICATION. The agent holds the context window open and polls for a response or establishes a websocket connection. This pattern is resource-intensive but ensures immediate consistency.
    \item Asynchronous Interrupt (Callback Mechanism). It is used for SOLICITATION or non-urgent inquiries. The agent serializes its current state (Check-pointing), sends the request, and suspends the process. Upon receiving a response (via a Webhook event), the agent "wakes up," restores the state, and injects the human input into the context window.
\end{enumerate}

By decoupling the communication logic into these patterns, A2H ensures that agents do not idle wastefully while waiting for human input, thereby optimizing computational resource utilization within the agent infrastructure.

\subsection{Unified Messaging Abstraction (UMA)}
The final component addresses the heterogeneity of communication channels. Agents typically output structured data (JSON), while humans communicate via natural language on platforms like Slack, Teams, or WeChat. The \textit{Unified Messaging Abstraction (UMA)} acts as a bidirectional translation layer.

\subsubsection{A2H-JSON Schema (Agent Output)}
We standardize the agent's output into a rigorous format called \textit{A2H-JSON}. This schema decouples the message content from the presentation layer. An A2H-JSON object consists of:
\begin{itemize}
\item \texttt{type}: The category of interaction (e.g., QUESTION, CONFIRMATION, ALERT).
\item \texttt{summary}: A one-line TL;DR of the issue.
\item \texttt{body}: Detailed context, supporting Markdown rendering.
\item \texttt{actions}: Structured options for the human (e.g., ["Approve", "Reject", "Modify"]).
\end{itemize}

\subsubsection{Channel Adapters (Rendering)}
The UMA layer employs Channel Adapters to render A2H-JSON into platform-specific UI elements.
\begin{itemize}
\item Slack/Teams: Renders as interactive "Block Kit" or "Adaptive Cards" with clickable buttons.
\item Email: Renders as HTML with deep links.
\item CLI: Renders as colored ASCII text.
\end{itemize}

\subsubsection{Response Normalization (Human Input)}
When a human replies (via text or button click), the Adapter normalizes the input back into a structured format for the agent. For instance, a button click "Approve" on Slack is translated into:

\begin{lstlisting}{json}
{
  "interaction_id": "uuid-1234",
  "human_id": "human://alice.eng",
  "decision": "APPROVED",
  "feedback": null
}
\end{lstlisting}

This ensures the agent can ingest human feedback deterministically as part of its observation space, closing the loop of the autonomous system.

\section{Case Study}
\begin{figure}[htbp]
    \centering
    \includegraphics[scale=0.37]{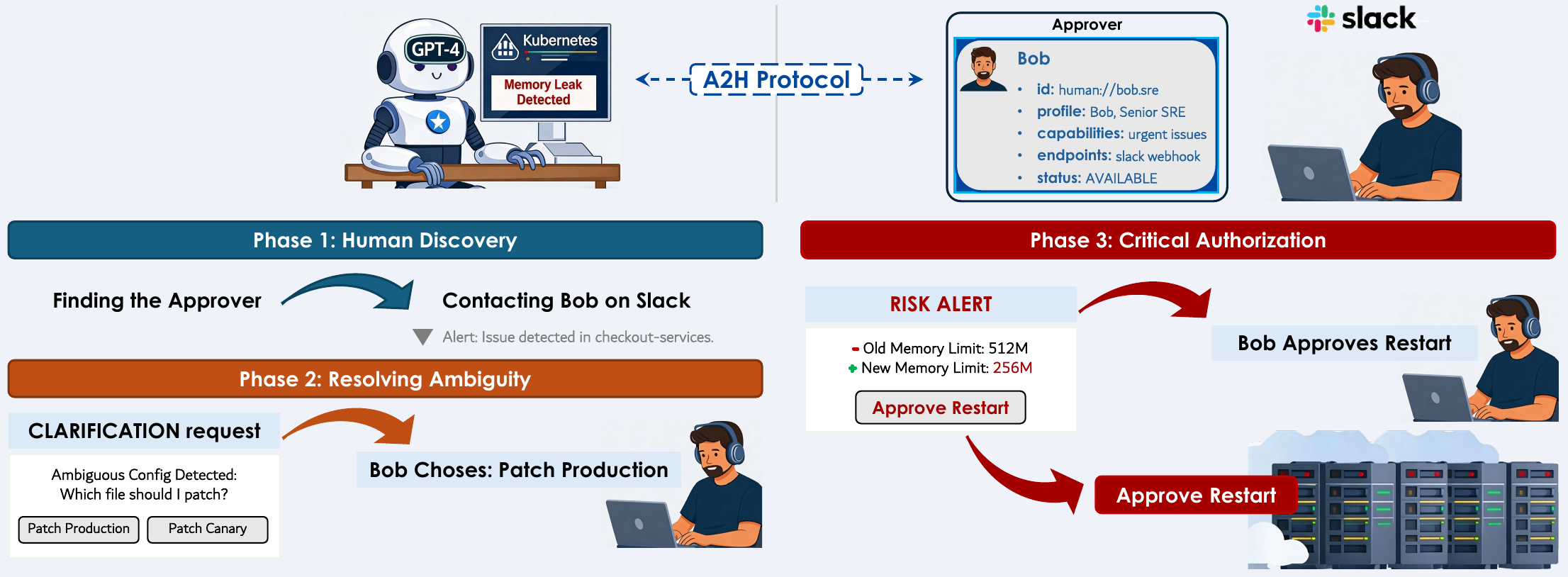}
    \caption{A2H case study.}
    \label{figure_case}
\end{figure}

To demonstrate the efficacy of the A2H protocol in a realistic environment, we deployed an A2H-enabled agent in a simulated DevOps scenario. The objective is to showcase the end-to-end workflow: from \textit{Human Discovery} (locating the expert) to \textit{Ambiguity Resolution} (clarification) and finally \textit{Critical Action Authorization} (permission).
The collaboration between an AI agent and a human is illustrated in Figure \ref{figure_case}. 

\subsection{Scenario Setup}
We define a task where an autonomous agent, powered by GPT-4, is responsible for monitoring and fixing a production service named checkout-service. 

\begin{enumerate}
    \item The Problem: The service experiences a memory leak.
    \item The Agent: A ReAct-based agent equipped with standard CLI tools (kubectl, git, logs) and the A2H Client.
    \item The Human: "Bob," a Senior SRE (Site Reliability Engineer), reachable via Slack.
    \item The Environment: The A2H Registry contains a Human Card for Bob with the tag ["sre", "kubernetes", "approver"].
\end{enumerate}

\subsection{Workflow Execution}
The execution trace follows three distinct phases, corresponding to the components defined in our Methodology.

\textbf{Phase 1: Human Discovery (Human Card)}

The agent analyzes the logs and identifies that the crash is related to a Kubernetes configuration. To proceed, the agent needs to identify the service owner. Instead of blindly messaging a general channel, the agent queries the A2H Registry:

\begin{equation}
    Find(h\in Registry \mid "kubernetes" \in h.S \land h.\delta=\text{AVAILABLE})
\end{equation}

\textbf{Result}: The registry returns Bob’s identity (human://bob.sre) and his preferred endpoint for urgent issues (slack\_webhook). This demonstrates the shift from broadcast communication to targeted addressing.

\textbf{Phase 2: Resolving Ambiguity (Clarification Trigger)}
The agent generates a fix but encounters two potential configuration files: deployment.yaml and deployment-canary.yaml. It cannot determine which file is currently active. The confidence score for editing either file falls below the threshold ($\tau_{amb} < 0.8$), triggering a CLARIFICATION request.

There are three steps:

\begin{enumerate}
    \item Agent Output (A2H-JSON): The agent constructs the following payload:
    \begin{lstlisting}{json}
{
  "target": "human://bob.sre",
  "type": "CLARIFICATION",
  "summary": "Ambiguous Configuration Target",
  "body": "I identified a memory limit issue.  Multiple config files detected. Which one should I patch?",
  "options": ["deployment.yaml (Production)",
              "deployment-canary.yaml (Canary)"]
}
    \end{lstlisting}
    \item UMA Rendering (Slack): The Unified Messaging Abstraction layer translates this JSON into a Slack "Block Kit" message. Bob sees a structured card with a concise summary and two clickable buttons: \textbf{[Patch Production]} and \textbf{[Patch Canary]}.
    \item Human Feedback: Bob clicks \textbf{[Patch Production]}. The A2H layer captures this interaction and returns a structured observation to the agent: {"selected\_option": "deployment.yaml (Production)"}. The agent proceeds to patch the file.
\end{enumerate}

\textbf{Phase 3: Critical Authorization (Permission Trigger)}
After applying the patch, the agent intends to restart the production cluster to apply changes. The protocol manifest flags (kubectl rollout restart) as a critical action, triggering the PERMISSION primitive.

Interaction as following:
\begin{enumerate}
    \item Blocking State: The agent enters a SUSPENDED state (Synchronous Blocking Pattern) to prevent accidental outages.
    \item Visualization: A "Risk Alert" card is sent to Bob, highlighting the diff of the changes and a red [Approve Restart] button.
    \item Resolution: Bob reviews the diff and clicks Approve. The agent receives the signal TRUE and executes the restart command.
\end{enumerate}

\subsection{Protocol Efficacy Analysis}
Table \ref{tab:case_study} compares the A2H-enabled workflow against a traditional Baseline Agent (which treats humans as simple chat interfaces).

\begin{table}[ht]
    \centering
    \caption{Comparison of A2H vs. Baseline Agent in DevOps Scenario}
    \label{tab:a2h_vs_baseline}
    \small
    \begin{tabularx}{\textwidth}{>{\RaggedRight\arraybackslash}p{2.5cm} >{\RaggedRight\arraybackslash}X >{\RaggedRight\arraybackslash}X}
        \toprule
        \textbf{Feature} & \textbf{Baseline Agent (Chat-based)} & \textbf{A2H-Enabled Agent} \\
        \midrule
        Addressing & 
        Manual (User must be in the chat loop) & 
        Dynamic Discovery (Finds expert via tags) \\
        
        Ambiguity & 
        Hallucinates a choice or loops & 
        Structured Solicitation (Clarification Trigger) \\
        
        Presentation & 
        Raw Text / JSON dumps & 
        Native UI Components (Buttons/Forms) \\
        
        Safety & 
        No formal guardrails & 
        Formal Permission Gates (Criticality Trigger) \\
        
        Result & 
        High risk of error or stall & 
        Successful, safe resolution \\
        \bottomrule
    \end{tabularx}
    \label{tab:case_study}
\end{table}

This case study confirms that A2H successfully bridges the gap between autonomous logic and human oversight, transforming complex decision points into standardized, safe interaction primitives.
\section{Conclusion}
In this paper, we presented A2H (Agent-to-Human), a novel protocol designed to bridge the gap between autonomous agent ecosystems and human participants. While existing protocols have successfully standardized agent-to-agent interactions, they have largely relegated humans to the role of external observers or manual triggers. A2H redefines this relationship by transitioning from a \textit{human-as-trigger} model to a \textit{human-as-node} architecture, treating humans as resolvable, discoverable, and addressable entities within the agent infrastructure.

We believe A2H represents a foundational step toward human-connected intelligent infrastructures, where AI agents and humans collaborate through standardized communication protocols.

\bibliographystyle{unsrt}
\bibliography{reference/references}

\end{document}